\documentstyle[prb,aps,multicol,epsf]{revtex}

\newlength{\colwidth}
\begin{document}
\draft
\title{Ultrasmall double junction in terms of orthogonal polynomials}
\author{Heinz--Olaf M{\"u}ller,\cite{email}}
\address{Department of Physics, Norwegian Institute of Technology,
Norwegian University of Science and Technology, N-7034 Trondheim,
Norway}
\author{Andreas H{\"a}dicke,}
\address{Institut f{\"u}r Festk{\"o}rperphysik,
  Friedrich--Schiller--Universit{\"a}t Jena, D-07740 Jena, F.\ R.\
  Germany} 
\author{Ulrik Hanke, and K. A. Chao}
\address{Department of Physics, Norwegian Institute of Technology,
Norwegian University of Science and Technology, N-7034 Trondheim,
Norway}

\maketitle

\begin{abstract}
  The ``orthodox theory'' of a single electron double junction is
  dealt with. It is shown that the stationary solution of the
  underlying master equation allows the construction of any
  time--dependent solution in terms of orthogonal polynomials. The
  approach pays off if the stationary solution becomes simple. Two
  special cases are considered. We use the time--dependent solution to
  calculate the current noise in these cases.
\end{abstract}

 \draft

\ifpreprintsty\relax\else\begin{multicols}{2}\fi

\section{Introduction}

Due to its simplicity the normal metal single electron double junction
(1E2J, shown schematically in Fig.~\ref{fig1}) is one of the
best--studied systems of single electronics, both experimentally and
theoretically. The common theoretical approach uses ``orthodox
theory''\cite{ave3,ave2} to find a master equation for the charge
probability of the island between the two junctions. The stationary
solution of this master equation is known
analytically.\cite{bur1,seu1,amm2} Since the derivation of the
equation itself restricts its solution to slow processes on the time
scale of $R_{\Sigma}C_{\Sigma}=(R_1+R_2)(C_1+C_2)$ the system appears
more or less transparent within ``orthodox theory.''

However, if one applies ``orthodox theory'' to the 1E2J one realizes
that the structure of the stationary solution restricts its practical
use to both low voltages and temperatures where the energy scale is
the Coulomb energy $E_{\rm C}=e^2/(2C_{\Sigma})$.  Therefore the
investigation of $n$--level systems, taking into account the most
important charge states only, was introduced.\cite{wan1} Using this
method, main features of single electron tunneling (SET), such as
Coulomb blockade or the Coulomb staircase, can be understood
theoretically.

Here we present an alternative method of solving the master equation
that is mainly applicable at either high voltage and negligible
temperature or high temperature and small voltage. In these cases the
stationary solution of the master equation simplifies and provides
access of the full time--dependent solution. In this regard the
presented approach is complementary to that of a $n$--level system. We
use this solution for the calculation of the current and the current
noise in a 1E2J for frequencies $f\ll(R_{\Sigma}C_{\Sigma})^{-1}$.

The physical relevance of our results lies in clarifying the
asymptotic behavior of the 1E2J rather than in unveiling new SET
effects. In fact, the latter can not be expected in the considered
domain. Furthermore, the treatment may be used as an easy--to--handle
approximation in specific cases.

In Sec.~\ref{main} we present the general approach. In
Secs.~\ref{ohmic} and~\ref{thermal} this general approach is applied
to the case of a high voltage at low temperature (ohmic limit) and low 
voltage but high temperature (thermal limit), respectively. Both
stationary mean current through a 1E2J and current noise power
spectrum are calculated. The results are discussed and limiting cases
are compared with known results.

\section{Method}
\label{main}

Let $r_{\mu}(x)$ and $l_{\mu}(x)$ denote the tunneling rates in right
and left direction (see Fig.~\ref{fig1}) across the $\mu$--th junction
if $x$ excess charges reside on the island between the junctions. Then
the matrix
\begin{eqnarray*}
  {\mathbf W}(y,x) & = & (-1)^{1+x+y}\big\{\big[r_1(x)+l_2(x)\big]
  \big[\delta_{y\,x+1}+\delta_{y\,x}\big]\\ 
  & & +\big[r_2(x)+l_1(x)\big]\big[\delta_{y\,x}+\delta_{y\,x+1}\big]
  \big\} 
\end{eqnarray*}
enables to write the ``orthodox'' master equation
\begin{equation}
\label{me}
|\dot{\sigma}\rangle = {\mathbf W}|\sigma\rangle,
\end{equation}
with the probability $\sigma(x)$ of $x$ excess charges on the island
as the $x$--th component of the vector $|\sigma\rangle$. Due to the
underlying tunneling of charges $x$ is integer. The left and right
eigenfunctions are given by 
\begin{eqnarray*}
  {\mathbf W}|\phi_n\rangle & = & -\lambda_n|\phi_n\rangle,\\ 
  \langle\psi_n|{\mathbf W} & = & -\lambda_n\langle\psi_n|,
\end{eqnarray*}
with the eigenvalues $\lambda_n\ge0$. It can be shown\cite{kam1} that
one of the eigenvalues equals zero (let it be $\lambda_0$).  The
corresponding eigenfunction $|\phi_0\rangle$ describes the stationary
solution of the master equation (\ref{me}). It obeys the detailed
balance condition
\begin{equation}
\label{db}
{\mathbf W}(x,x+1)\phi_0(x+1)={\mathbf W}(x+1,x)\phi_0(x)
\end{equation}
Left and right eigenfunctions are related through\cite{kam1}
\begin{equation}
\label{lr}
\phi_n(x)=\psi_n(x)\phi_0(x).
\end{equation}
From Eq.~(\ref{lr}) we see that $\psi_0(x)=1$. The orthogonality
relation of the left eigenfunctions reads
\begin{equation}
\label{ortho}
\sum_x\psi_m(x)\psi_n(x)\phi_0(x) = \delta_{mn}.
\end{equation}
Thus, the eigenfunctions $\psi_n(x)$ form a set of orthogonal
polynomials with the weight function $\phi_0(x)$ (``SET
polynomials''). As an example, these polynomials are shown in
Table~\ref{tab1} for a special case. In general, $\psi_n(x)$ is a
polynomial of order $n$ in $x$.

If the stationary solution of the master equation (\ref{me}) is known,
an arbitrary time dependent one follows readily from Eqs.~(\ref{lr})
and (\ref{ortho}):
\begin{displaymath}
  |\sigma(t)\rangle = \sum_n b_n{\rm e}^{-\lambda_n(t-t_0)}|\phi_n\rangle,
\end{displaymath}
where the $b_n$ are adjusted to meet the initial probability
distribution $|\sigma(t_0)\rangle$, $b_n=\langle\psi_n|\sigma(t_0)\rangle$.
The eigenvalues $\lambda_n$, which determine the time--dependence,
result from Eq.~(\ref{me}):
\begin{eqnarray}
\label{lambda}
\lambda_n & = & {\mathbf
  W}(x-1,x)\bigg[1-\frac{\psi_n(x-1)}{\psi_n(x)}\bigg]\\ 
	& & +{\mathbf W}(x+1,x)\bigg[1-\frac{\psi_n(x+1)}{\psi_n(x)}\bigg],
\qquad\forall x.\nonumber
\end{eqnarray}
In summary, the stationary solution of the master equation
complemented with an initial condition determines the full
time--dependent solution.

\section{Ohmic limit}
\label{ohmic}

Now we turn to the case of zero temperature. Then the detailed balance
property (\ref{db}) causes the ratio $\phi_0(x+1)/\phi_0(x)$ to be a
rational function of the bias voltage $v=V\,C_{\Sigma}/e$, and of the
dimensionless charges $q=Q/e$ and $x$. Whereas $x$ describes the
integer part of the charge, which is altered during tunneling
processes, the parameter $q$ expresses the influence of background
charges and/or an applied gate voltage. In the following we will make
use of the additional dimensionless parameters
$\kappa_{\mu}=C_{3-\mu}/C_{\Sigma}$ and
$\varrho_{\mu}=R_{\mu}/R_{\Sigma}$, $\mu=1,2$. If $v$ is integer and
obeys
\begin{equation}
\label{cond}
\kappa_2v+q-1/2=m-0^+
\end{equation} 
for an arbitrary other integer $m$, the ratio $\phi_0(x+1)/\phi_0(x)$
can be written as
\begin{displaymath}
  \frac{\phi_0(x+1)}{\phi_0(x)} = \frac{\varrho_2}{\varrho_1}\,
  \frac{N-k}{k+1},
\end{displaymath}
with $N=v-1$ and $k=x-x_{\min}=x+m-1$. $x_{\rm min}$ denotes the
lowest of all accessible charge states, $x_{\rm min} =
-[\kappa_2v+q+1/2]$, and $[\ldots]$ the integer part of $\ldots$.
Therefore $\phi_0(x)$ is a binomial distribution in these cases:
\begin{displaymath}
  \phi_0(x)= {N \choose k}\varrho_2^{k}\varrho_1^{N-k}.
\end{displaymath}
The given condition (\ref{cond}) defines a number of points in the
$V$--$Q$ plane, as shown in Fig.~\ref{fig2} (``binomial points'').
However, if $1\ll k\ll N$, the condition (\ref{cond}) is fulfilled for
all $v$ and $q$. This is the case for high bias voltage $V$.

For the binomial distribution the polynomials $\psi_n(x)$ are known.
They can be expressed in terms of Krawtchouk polynomials\cite{koe1} as
follows:
\begin{displaymath}
  \psi_n(x)=K_n\big(x-x_{\min};\varrho_2,N\big) \sqrt{{N\choose n}
    \bigg(\frac{\varrho_2}{\varrho_1}\bigg)^n}.
\end{displaymath}
The eigenvalues are determined from Eq.~(\ref{lambda}), which is the
difference equation (1.10.4) of Ref.~\onlinecite{koe1}, yielding
\begin{equation}
  \label{lambdat0}
  \lambda_n =
  \frac{n}{R_{\Sigma}C_{\Sigma}}\,\frac{1}{\varrho_1\varrho_2} =
  \frac{n}{R_{\|}C_{\Sigma}}
\end{equation}
with $R_{\|}=R_1R_2/R_{\Sigma}$.

The calculation of the stationary current follows the standard
procedure\cite{ave3,ave2} and yields
\begin{equation}
  \label{it0}
  \langle I\rangle = \frac{e(v-1)}{R_{\Sigma}C_{\Sigma}}.
\end{equation}
One verifies that this formula describes the asymptotic behavior for
large bias correctly. For lower bias the ``binomial points'' of
Fig.~\ref{fig2} correspond to voltages where the $\langle
I\rangle$--$V$ characteristic of the 1E2J touches the asymptotic line.
For the symmetric 1E2J with $q=0$ this happens for $v=2\,m+1$ with
integer $m$.

Next we present our result of the current noise of a 1E2J. The method
of calculation has been described in Refs.~\onlinecite{kor2}
and~\onlinecite{han3}. It takes advantage of the representation of the
conditional probability $p(m,t;m',t')$ and the matrix elements of
${\mathbf W}$ in terms of the eigenfunctions $\psi_n(m)$. After
several steps (see Appendix) we obtain
\begin{equation}
  \label{st0}
  S(0) = 2\,e\langle I\rangle\big(1-2\varrho_1\varrho_2\big)
\end{equation}
for zero frequency. Since a realistic estimation shows that
$(R_{\Sigma}C_{\Sigma})^{-1}\ge1{\rm GHz}$ typically, the result
(\ref{st0}) is valid in a wide range of frequencies, down to the
regime where $1/f$--noise becomes predominant.

Eq.~(\ref{st0}) yields the ordinary shot noise formula in the limit of
an extremely asymmetric 1E2J ($\varrho_1\to1\wedge\varrho_2\to0$ or vice
versa), where the current across both junctions is uncorrelated. The
shot noise is halved for a symmetric 1E2J in agreement with earlier
results.\cite{kre3,han2}

Additional noise suppression in double junctions in dependence on the
bias voltage was both theoretically predicted\cite{kre3,han2} and
experimentally observed.\cite{bir1} Our result (\ref{st0}) does not
recover this suppression since the ``binomial points'' are located at
bias values where this suppression does not occur.

In Fig.~\ref{fig3} and Fig.~\ref{fig4} the results for symmetric and
asymmetric 1E2J are shown. For the symmetric system the noise is
suppressed. An additional suppression occurs between the ``binomial
points'' in comparison to $2\,e\langle I\rangle$. For higher bias the
use of (\ref{st0}) as approximation improves. This improvement is best
for the symmetric 1E2J.

\section{Thermal limit}
\label{thermal}

Let us now consider the case of dominating tem\-pe\-rature $k_{\rm
  B}T\equiv1/\beta\ge E_{\rm C}$. The equilibrium distribution is then
a Gaussian with mean $(\kappa_1-\varrho_1)v$ and width $(2\beta E_{\rm
  C})^{-1/2}$
\begin{equation}
  \label{phi0th}
  \phi_0(x) = \sqrt{\frac{\beta E_{\rm C}}{\pi}}
  \exp\big(-[x-(\kappa_1-\varrho_1)v]^2\beta E_{\rm C}\big),
\end{equation}
which follows from detailed balance (\ref{db}), or equivalently, from
a thermodynamic consideration. In order to determine the eigenvalues,
we consider Eq.~(\ref{lambda}) in the limit $x\to\pm\infty$, where it
yields again the temperature independent values of
Eq.~(\ref{lambdat0}).  While our consideration has been exact within
orthodox theory so far, the treatment requires several approximations
from now on. Since many states are significantly occupied, the above
charge state sums can be replaced by integrals $\sum_x\to\int{\rm
  d}x$. Then, the SET polynomials are given in terms of Hermite
polynomials
\begin{equation}
  \label{psinth}
  \psi_n(x) = \frac{H_n\big([x-(\kappa_1-\varrho_1)v]\sqrt{\beta
      E_{\rm C}}\big)}{\sqrt{2^nn!}}.
\end{equation}

For the calculation of the stationary current let us introduce a
current vector $\langle\iota|$ by
\begin{eqnarray*}
  \langle I\rangle & = & \langle\iota|\sigma\rangle\\
  \iota(x) & = &
  \sum_{\mu=1}^2\kappa_{\mu}\big[r_{\mu}(x)-l_{\mu}(x)\big]\\
  \langle\iota| &  = & \sum_na_n\langle\psi_n|,
\end{eqnarray*}
where a Taylor expansion in $(\beta E_{\rm C})$ is used to approximate
the high temperature behavior of $\langle\iota|$. Even if this
expansion yields positive powers in $x$, which are integrated later
on, its use is justified by $\phi_0(x)$ of (\ref{phi0th}), which
decreases very rapidly with increasing $|x|$. For the stationary
current we find
\begin{displaymath}
  \langle I\rangle = \langle\iota|\phi_0\rangle = a_0
  \stackrel{\beta\to0}{\longrightarrow} \frac{V}{R_{\Sigma}}
  \left(1-\frac{\beta E_{\rm C}}{3}\right).
\end{displaymath}
This high temperature result is independent of the double junction's
symmetry. It agrees with the known result from SET
thermometry.\cite{pek1}

The calculation of the noise follows again
Refs.~\onlinecite{kor2,han3}. In terms of (\ref{phi0th}) we use
\begin{displaymath}
  \phi_0(x\pm1) = \phi_0(x)\,{\rm e}^{-\beta E_{\rm
    C}\{1\pm2[x-(\kappa_1-\varrho_1)v]\}}. 
\end{displaymath}
Furthermore, high temperature expansions are used again to simplify
the calculation which is lengthy, but straightforward. Restricting to
the leading term and its first order correction results in
contributions from the eigenvalues $\lambda_{0,1}$ only. After
carefully collecting all relevant contributions the zero frequency
noise power
\begin{equation}
  \label{stth}
  S(0) = \frac{4k_{\rm B}T}{R_{\Sigma}}\,
  \bigg[1-\frac{\beta E_{\rm
      C}}{3}\bigg(2-\frac{\kappa_1^2}{\varrho_1}
  -\frac{\kappa_2^2}{\varrho_2}\bigg)\bigg]
\end{equation}
is found. This result coincides with the high temperature limit of
Ref.~\onlinecite{kor2}, but gives a correction to the known expression
as well. In case of symmetric 1E2J (or more generally: for
$\kappa_i=\varrho_i$) the term $(\ldots)$ in (\ref{stth})
simplifies to unity and it becomes clear that the Nyquist formula
yields an upper bound of the noise of a 1E2J. This fact expresses the
partial coherence between tunneling events on both junctions. At high
temperature, however, this coherence get lost. This behavior does not
correspond to the $\coth$--formula\cite{bir1,zie1}
\begin{displaymath}
  S(0) = 2e\langle I\rangle\coth\bigg(\frac{e\,V}{2k_{\rm B}T}\bigg),
\end{displaymath}
using the current $I$ through the system and the bias voltage $V$,
which predicts elevated noise above $4k_{\rm B}T/R_{\Sigma}$. We want
to point out that the $\coth$--formula was derived for the single
junction (1E1J) and does not allow for coherence effects of both
junctions.

\section{Conclusions}

We have presented a method allowing access to the time--dependent
solution of the master equation that is used in the semiclassical
``orthodox theory'' to describe the behavior of an ultrasmall single
electron double junction. This method is based on general properties
of master equations. We can show that the stationary solution of the
master equation is sufficient to determine any time--dependent
solution by quadratures on the range of allowed charge states. For
special cases of the stationary solution this procedure simplifies to
an approximatively analytically solvable problem. Two of these cases are
presented in this paper in detail. After calculating the respective
mean currents the time--dependent solution is used to find expressions
for the current noise power spectra. The obtained results are
discussed. It turns out that the presented method together with the
usual $n$--level approach enable access to time--dependent solution in
a wide parameter range of the bias voltage $V$ and the temperature
$T$.

\acknowledgments

Financial support by DAAD (HOM), and NFR (UH) is gratefully
acknowledged. For stimulating discussions we like to thank
J.~A.~Melsen and A.~Matulis.

\appendix

\section{Calculation for the ohmic limit}

The eigenvalues (\ref{lambdat0}) follow directly from the difference
equation (1.10.4) in Ref.~\onlinecite{koe1}. In the 
evaluation of the noise we follow strictly the description of
Ref.~\onlinecite{han3}, namely
\begin{equation}
  \eqnum{A1}\label{a1}
  S^{\rm Sh}_{\mu} =
e^2\sum_n\big[r_{\mu}(n)+l_{\mu}(n)\big]\phi_0(n),
\end{equation}

\ifpreprintsty\relax\else\end{multicols}\noindent\rule{\colwidth}{0.4pt}\fi

\begin{eqnarray}
  \eqnum{A2}\label{a2}
S^{\rm c}_{\mu\mu'}(t-t') & = &
e^2\sum_{m\,m'}\big[r_{\mu}(m)-l_{\mu}(m)\big]p(m,t;m',t')\\
	& & \times\big[r_1(m'-1)\phi_0(m'-1)-l_1(m'+1)\phi_0(m'+1)
	+r_2(m'+1)\phi_0(m'+1)-l_2(m'-1)\phi_0(m'-1)\big],\nonumber
\end{eqnarray}

\ifpreprintsty\relax\else\hfill\rule{\colwidth}{0.4pt}\begin{multicols}{2}\fi

\begin{eqnarray}
  \eqnum{A3}\label{a3}
  S_{\mu\mu'}(t) & = & -\langle I\rangle^2+
	\delta(t)\delta_{\mu\mu'}S^{\rm Sh}_{\mu}+S^{\rm
	c}_{\mu\mu'}(t),\\
  \eqnum{A4}
  S(t) & = & \sum_{\mu\mu'}\kappa_{\mu}\kappa_{\mu'}S_{\mu\mu'}(t),\\
  \eqnum{A5}\label{a5}
  S(\omega) & = & 4\int^{\infty}_0\!\!\!{\rm d}t\cos(\omega t)S(t) 
\end{eqnarray}
is finally the noise power spectrum of the current that is considered
in the zero frequency limit $\omega\to0$. 

The conditional probability $p(m,t;m',t')$ can be expressed in terms
of the eigenfunctions of ${\mathbf W}$ 
\begin{displaymath}
  p(m,t;m',t') = \sum_{n=0}^{N}\psi_n(m')\phi_n(m) {\rm
    e}^{-\lambda_n(t-t')}.
\end{displaymath}
From the orthogonality relation\cite{kam1}
\begin{displaymath}
  \sum_{n=0}^{N}\psi_n(m')\phi_n(m)=\delta_{m\,m'}
\end{displaymath}
we see that the correct initial condition of $p(m,t;m',t')$ is obeyed
at $t=t'$. Now the different noise contributions are calculated. Due
to the vanishing temperature one finds $S_{\mu}^{\rm Sh}=e\langle
I\rangle$ independent of the junction number $\mu$ and the system's
symmetry ($\kappa_{1,2}$, $\varrho_{1,2}$). The non-diagonal terms of 
$S_{\mu\mu'}^{\rm c}$ simplify using the detailed balance property
(\ref{db}). With the rates $r_{\mu}(m)$ (the current direction
has to be considered only) $S_{\mu\mu'}^{\rm c}$ is transformed into
\begin{eqnarray*}
  S_{\mu\mu'}^{\rm c}(t-t') & = & e^2\sum_{n=0}^{N} {\rm
    e}^{-\lambda_n(t-t')} \times\\ & &
  \sum_{m\,m'}r_{\mu}(m)\phi_n(m)r_{3-\mu'}(m')\phi_n(m').
\end{eqnarray*}
The term corresponding to $n=0$ in the last equation results in
$\langle I\rangle^2$, which cancels against the same term in the 
noise expression (\ref{a3}). The rates $r_{\mu}(m)$ are linear
functions of their arguments. Hence, they can be expanded into the
left eigenfunctions $\psi_n(m)$ with contributions of $\psi_{0,1}(m)$
only. Note, that the $\Theta$--function of the rates can be omitted in
the considered ``binomial points''. Using the representation
$r_{\mu}(m) = c_{\mu\,0}\psi_0(m)+c_{\mu\,1}\psi_1(m)$ the expression
above is further simplified to
\begin{displaymath}
  S_{\mu\mu'}^{\rm c}(t-t') =
  e^2c_{\mu\,1}c_{3-\mu'\,1}\exp[-\lambda_1(t-t')].
\end{displaymath}
In detail, we find
\begin{eqnarray*}
  c_{\mu\,0} & = & \langle I\rangle/e,\\ c_{\mu\,1} & = &
  \frac{(-1)^{\mu+1}}{R_{\Sigma}C_{\Sigma}}
  \sqrt{N\frac{\varrho_1\varrho_2}{\varrho_{\mu}^2}},
\end{eqnarray*}
using the self--duality property of the Krawtchouk polynomials.
Finally, assemblage of the noise follows the line of
Ref.~\onlinecite{han3} straightforwardly. For zero frequency we obtain
\begin{displaymath}
  S(0) = 2e\langle I\rangle(1-2\varrho_1\varrho_2) =
  \frac{2e^2(v-1)}{R_{\Sigma}C_{\Sigma}}(1-2\varrho_1\varrho_2),
\end{displaymath}
where the stationary current is expressed via (\ref{it0}).

\ifpreprintsty\relax\else\appendix\stepcounter{section}\fi

\section{Calculation for the thermal limit}

For the calculation of the eigenvalues $\lambda_n$ the approximations
\begin{eqnarray*}
  {\mathbf W}(x-1,x) = r_2(x)+l_1(x) 
  & \approx & \frac{1}{R_{\Sigma}C_{\Sigma}}\,\frac{x}{\varrho_1\varrho_2}\\
  1-\frac{\psi_n(x-1)}{\psi_n(x)} & \approx & \frac{n}{x}\\
  {\mathbf W}(x+1,x) = r_1(x)+l_2(x) & \approx & 0
\end{eqnarray*}
are found for $x\gg1$. Introducing into (\ref{lambda}) results in the
eigenvalues (\ref{lambdat0}). A similar consideration for $x\ll-1$
yields the same result.

The calculation of the mean current starts from the Gaussian
approximation of the stationary solution $\phi_0(x)$ of (\ref{phi0th})
and the corresponding left eigenfunctions (\ref{psinth}). The relevant
expansion of the current operator $\iota(x,v)$ is simplified by the
use of the shifted charge $y=[x-(\kappa_1-\varrho_1)v]$:
\begin{displaymath}
\iota(x,v) \stackrel{\beta\to0}{\longrightarrow} 
\bigg(\frac{V}{R_{\Sigma}}
-\frac{e(\kappa_1-\varrho_1)y}{R_{||}C_{\Sigma}}\bigg)
\bigg(1-\frac{\beta E_{\rm C}}{3}\bigg).
\end{displaymath}
The determination of $a_i$ results in contributions to $a_{0,1}$, in
detail
\begin{eqnarray*}
  a_0 & = & \frac{V}{R_{\Sigma}}\bigg(1-\frac{\beta E_{\rm
      C}}{3}\bigg)\\
  a_1 & = & \frac{e(\kappa_1-\varrho_1)}{R_{||}C_{\Sigma}\sqrt{2\beta
      E_{\rm C}}}\bigg(1-\frac{\beta E_{\rm C}}{3}\bigg).
\end{eqnarray*}
The calculation of the stationary current makes use of $a_0$ only.

The calculation of the noise follows a similar, though more
complicated, line of the current calculation above in the evaluation
of the expression (\ref{a1})--(\ref{a5}). By use of
(\ref{phi0th}) we find
\begin{eqnarray*}
  r_{1,2}(x\mp1)\phi_0(x\mp1) & = & l_{1,2}(x)\phi_0(x)\exp(2\beta
  E_{\rm C}\varrho_{1,2}v)\\
  l_{1,2}(x\pm1)\phi_0(x\pm1) & = & r_{1,2}(x)\phi_0(x)\exp(-2\beta
  E_{\rm C}\varrho_{1,2}v).
\end{eqnarray*}
Furthermore, the abbreviations
\begin{eqnarray*}
  i_{\mu}(x) & = & e[r_{\mu}(x)-l_{\mu}(x)]\\
  j_{\mu}(x) & = & e[l_{\mu}(x)\exp(2\beta E_{\rm C}\varrho_{\mu}v)-
  r_{\mu}(x)\exp(-2\beta E_{\rm C}\varrho_{\mu}v)]\\
  k_{\mu}(x) & = & e[r_{\mu}(x)+l_{\mu}(x)],
\end{eqnarray*}
allow the transformation of the correlation function into
\begin{equation}
  \eqnum{B2}
  S^{\rm c}_{\mu\mu'}(t-t') = \sum_n\exp[-\lambda_n(t-t')]
  \langle i_{\mu}|\phi_n\rangle\,\langle j_{\mu'}|\phi_n\rangle
\end{equation}
and the Shottky value is expressed as $S^{\rm Sh}_{\mu} = e\langle
k_{\mu}|\phi_0\rangle$.  The representation of the relevant high
temperature expansion of $i_{\mu}$, $j_{\mu}$, and $k_{\mu}$ in terms
of the left eigenfunctions $\psi_n$ reads
\begin{eqnarray*}
  \langle i_{\mu}| & \approx & \frac{2E_{\rm C}}{e\,R_{\Sigma}}v
  \langle\psi_0|
  \mp\frac{2E_{\rm C}}{e\,R_{\Sigma}}\,\frac{1}{\varrho_{\mu}
    \sqrt{2\beta E_{\rm C}}}\bigg(1-\frac{\beta E_{\rm C}}{3}\bigg)
    \langle\psi_1|\\
  \langle j_{\mu}| & \approx & \frac{2E_{\rm C}}{e\,R_{\Sigma}}v
  \langle\psi_0|
  \pm\frac{2E_{\rm C}}{e\,R_{\Sigma}}\,\frac{1}{\varrho_{\mu}
    \sqrt{2\beta E_{\rm C}}}\bigg(1-\frac{\beta E_{\rm C}}{3}\bigg)
    \langle\psi_1|\\
  \langle k_{\mu}| & \approx & \frac{2}{e\,R_{\Sigma}\varrho_{\mu}}
  \bigg(\frac{1}{\beta}-\frac{E_{\rm C}}{3}\bigg)\langle\psi_0|
  +\frac{E_{\rm C}\sqrt{2}}{3e\,R_{\Sigma}\varrho_{\mu}}\langle\psi_2|.
\end{eqnarray*}
The calculation of the zero frequency noise power follows directly
(\ref{a3})--(\ref{a5}). Keeping the leading two orders in $\beta
E_{\rm C}$ only results in the expression (\ref{stth}) given in the
text.


\begin{references}

\bibitem{email}
hom@phys.unit.no.

\bibitem{ave3}
D.~V. Averin and K.~K. Likharev,  in {\em Mesoscopic Phenomena in Solids}, {\em
  Modern Problems in Condensed Matter Sciences (30)}, edited by B.~L.
  Altshuler, P.~A. Lee, and R.~A. Webb (Elsevier, Amsterdam, 1991), pp.\
  173--271.

\bibitem{ave2}
D.~V. Averin and K.~K. Likharev, Zh. Eksp. Teor. Fiz. {\bf 90},  733  (1986),
  (in Russian) [{\it Sov.\ Phys.--JETP} {\bf 63}, 427 (1986)].

\bibitem{bur1}
R.~E. Burgess, Physica {\bf 20},  1007  (1954).

\bibitem{seu1}
F. Seume and W. Krech, Ann. Phys. {\bf 1},  198  (1992).

\bibitem{amm2}
M. Amman {\it et~al.}, Phys. Rev. B {\bf 43},  1146  (1991).

\bibitem{wan1}
J.~C. Wan {\it et~al.}, Phys. Rev. B {\bf 43},  9381  (1991).

\bibitem{kam1}
N.~G. van Kampen, {\em Stochastic Processes in Physics and Chemistry}
  (Elsevier, Amsterdam, 1981).

\bibitem{koe1}
R. Koekoek and R.~F. Swarttouw, Technical Report No.~94-05, TU Delft,
  http://www.can.nl:80/~renes/new/, 
  ftp://ftp.twi.tudelft.nl/pub/publications/tech-reports\\
  /1994/DUT-TWI-94-05.ps.gz (unpublished).

\bibitem{kor2}
A.~N. Korotkov, Phys. Rev. B {\bf 49},  10381  (1994).

\bibitem{han3}
U. Hanke {\it et~al.}, Phys. Rev. B {\bf 51},  9084  (1995).

\bibitem{kre3}
W. Krech, A. H{\"a}dicke, and H.-O. M{\"u}ller, Int. J. Mod. Phys. B {\bf 6},
  3555  (1992).

\bibitem{han2}
U. Hanke, Y.~M. Galperin, K.~A. Chao, and N. Zou, Phys. Rev. B {\bf 48},  17209
   (1993).

\bibitem{bir1}
H. Birk, M.~J.~M. de~Jong, and C. Sch{\"o}nenberger, Phys. Rev. Lett. {\bf 75},
   1610  (1995).

\bibitem{pek1}
J. Pekola, K. Hirvi, J. Kauppinen, and M. Paalanen, Phys. Rev. Lett. {\bf 73},
  2903  (1994).

\bibitem{zie1}
A. van~der Ziel, {\em Noise in Solid State Devices and Circuits} (Wiley, New
  York, 1986).

\end{references}

\narrowtext
\begin{figure}
  \ifx\epsfxsize\undefined\relax\else
  \epsfxsize=\colwidth\epsffile{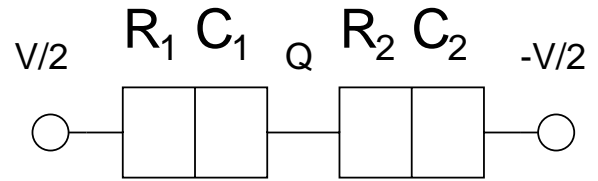}\fi
\caption{Circuit diagram of the ultrasmall double--tunnel junction,
  consisting of two tunnel junctions connected in series (with
  resistances $R_{1,\,2}$ and capacitances $C_{1,\,2}$, respectively)
  and an excess charge $Q$ on the island.}
\label{fig1}
\end{figure}

\begin{figure}
  \ifx\epsfxsize\undefined\relax\else
  \epsfxsize=\colwidth\epsffile{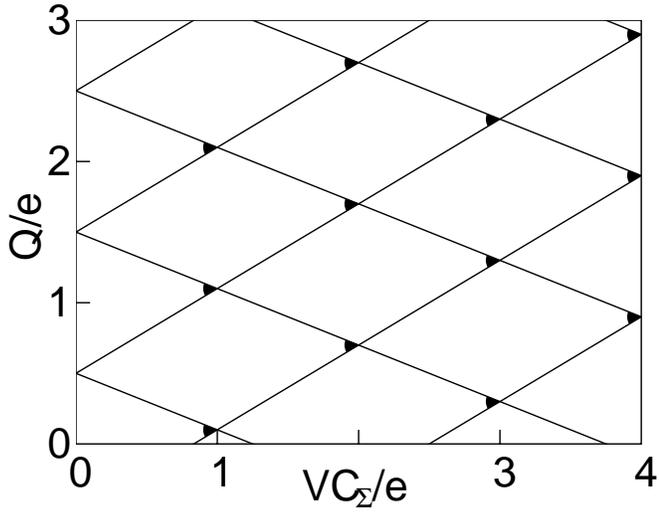}\fi
\caption{$V$--$Q$ plane for $C_2/C_1=1.5$. The tiles correspond to
  regions of fixed $N$ and the black corners indicate the points where
  the stationary charge distribution is binomial.}
\label{fig2}
\end{figure}

\begin{figure}
  \ifx\epsfxsize\undefined\relax\else
  \epsfxsize=\colwidth\epsffile{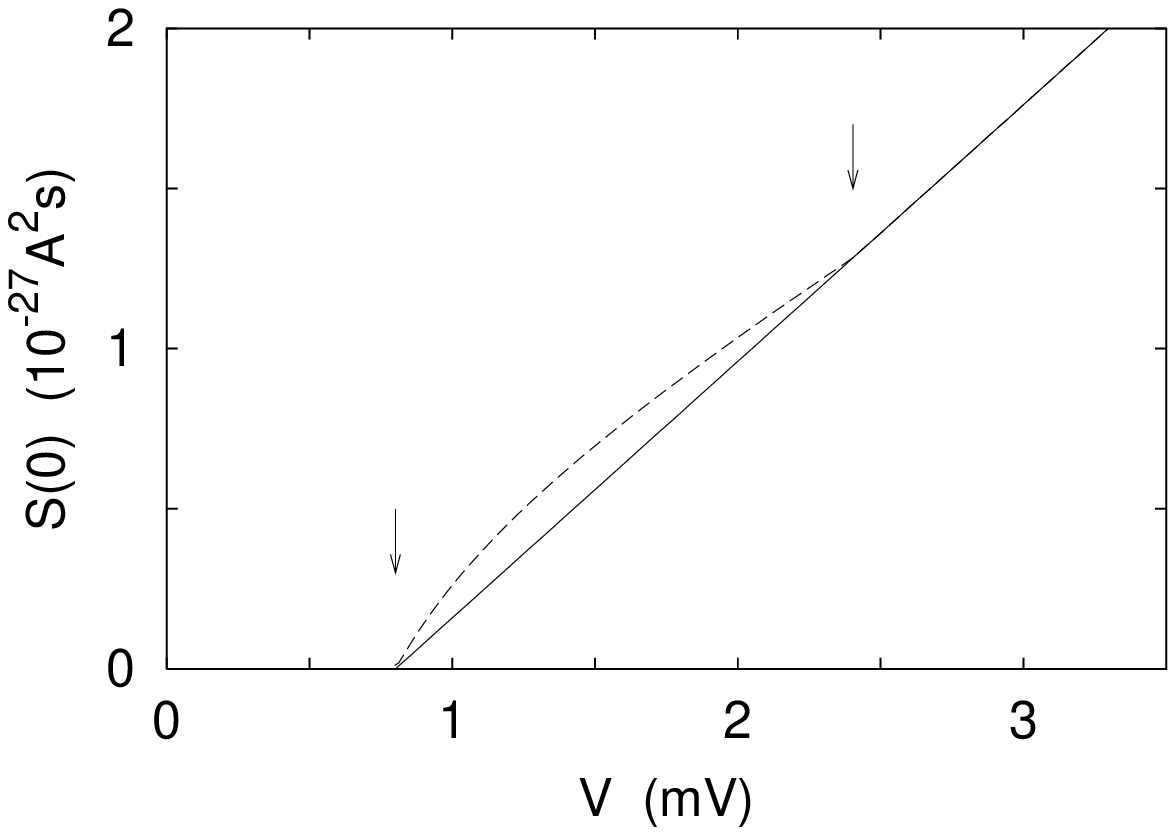}\fi
\caption{Noise in a symmetric 1E2J ($R_{\Sigma}=200{\rm k}\Omega$,
  $C_{\Sigma}=0.2{\rm fF}$). The full line represents Eq.~(\ref{st0})
  and the dashed line displays the $2\,e\langle I\rangle$. The arrows
  indicate the ``binomial points'' for this system.}
\label{fig3}
\end{figure}

\begin{figure}
  \ifx\epsfxsize\undefined\relax\else
  \epsfxsize=\colwidth\epsffile{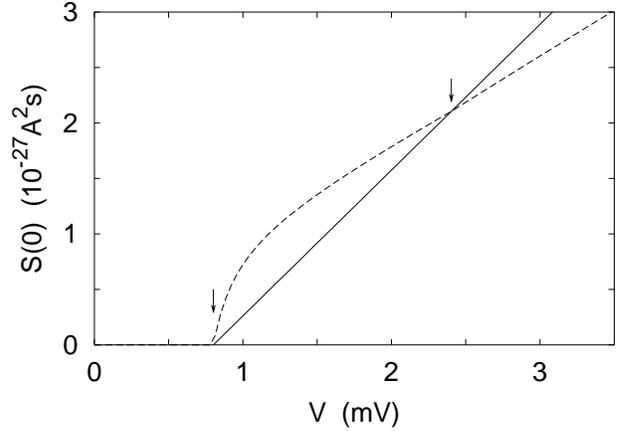}\fi
\caption{Noise in an asymmetric ($\kappa_1=\varrho_1=0.9$) 1E2J 
  ($R_{\Sigma}=200{\rm k}\Omega$, $C_{\Sigma}=0.2{\rm fF}$). The full
  line represents Eq.~(\ref{st0}) and the dashed line displays the
  $2\,e\langle I\rangle$. The arrows indicate the ``binomial points''
  for this system.} 
\label{fig4}
\end{figure}

\begin{table}
\begin{tabular}{lccc}
  & $c_{n0}$ & $c_{n1}$ & $c_{n2}$\\ \hline $n=0$ & $1$ & & \\ $n=1$ &
  $0$ & ${\displaystyle-\sqrt{\frac{3v-1}{2v-2}}}$ & \\ $n=2$ &
  $\displaystyle{-\sqrt{2\frac{v-1}{v+1}}}$ & $0$ &
  $\displaystyle{\frac{3v-1}{\sqrt{2v^2-2}}}$
\end{tabular}\medskip
\caption{Coefficients $c_{nm}$ of the SET polynomials for a symmetric
  1E2J with $Q=T=0$ at dimensionless voltages $v$ in units of
  $e/C_{\Sigma}$ so that $v\in(1;3]$. The polynomials are built from
  these coefficients via $\psi_n(x)=\sum_{m} c_{nm}x^m$.}
\label{tab1}
\end{table}

\ifpreprintsty\relax\else\end{multicols}\fi

\end{document}